\documentclass{IEEEtran}

\usepackage{graphicx}
\usepackage{amsfonts}
\usepackage{float}
\usepackage{bm}
\usepackage{algorithm}
\usepackage{algorithmic}
\usepackage{amsmath}
\usepackage{amsthm}

\usepackage[normalem]{ulem}                       
\newcommand\redout{\bgroup\markoverwith{\textcolor{red}{\rule[0.5ex]{2pt}{0.8pt}}}\ULon}

\usepackage[capitalise]{cleveref}                 
\crefname{equation}{}{}                           
\Crefname{equation}{}{}                           



\newcommand{\Tscr}{\mathcal T}

\newcommand{\matst}[1]{{\fontfamily{qcr}\selectfont #1}}

\newcommand{\dsum}{\displaystyle\sum\limits}

\newcommand{\bbR}{\mathbb R}

    \newcommand{\floor}[1]{\left\lfloor{#1}\right\rfloor}

    \renewcommand{\v}[1]{{\bf{#1}}}

\title{
    Exploiting flow dynamics for super-resolution in contrast-enhanced ultrasound
}

\author{Oren Solomon,~\IEEEmembership{Student Member,~IEEE,} Ruud J.G. van Sloun,~\IEEEmembership{Member,~IEEE}, Hessel Wijkstra, Massimo Mischi~\IEEEmembership{Senior member,~IEEE} 
\thanks{O. Solomon and R. J. G. van Sloun contributed equally to this work.}
and Yonina C. Eldar,~\IEEEmembership{Fellow,~IEEE} 
    \thanks{This project has received funding from the European Union's Horizon 2020 research and innovation program under grant agreement No. 646804-ERC-COG-BNYQ and starting grant No. 280209, and from the Ollendorf Foundation.}
\thanks{O. Solomon (e-mail:  orensol@tx.technion.ac.il) and Y. C. Eldar (e-mail: yonina@ee.technion.ac.il) are with the Department of Electrical Engineering, Technion—Israel Institute of Technology, Haifa 32000.}
\thanks{R. J. G. van Sloun (e-mail:  R.J.G.v.Sloun@tue.nl) and M. Mischi (e-mail: M.Mischi@tue.nl) are with the Department of Electrical Engineering, Eindhoven University of Technology, Eindhoven, The Netherlands.}
\thanks{H. Wijkstra (e-mail:  H.Wijkstra@tue.nl) is with the Department of Electrical Engineering, Eindhoven University of Technology, Eindhoven, The Netherlands and with the Department of Urology, Academic Medical Center - University of Amsterdam, Amsterdam, The Netherlands.}
}

\begin{document}
\maketitle
\begin{abstract}
Ultrasound localization microscopy offers new radiation-free diagnostic tools for vascular imaging deep within the tissue. Sequential localization of echoes returned from inert microbubbles with low-concentration within the bloodstream reveal the vasculature with capillary resolution. Despite its high spatial resolution, low microbubble concentrations dictate the acquisition of tens of thousands of images, over the course of several seconds to tens of seconds, to produce a single super-resolved image. 
Such long acquisition times and stringent constraints on microbubble concentration are undesirable in many clinical scenarios. To address these restrictions, sparsity-based approaches have recently been developed. These methods reduce the total acquisition time dramatically, while maintaining good spatial resolution in settings with considerable microbubble overlap. 
Here, we further improve sparsity-based super-resolution ultrasound imaging by exploiting the inherent flow of microbubbles and utilize their motion kinematics. While doing so, we also provide quantitative measurements of microbubble velocities. Our method relies on simultaneous tracking and super-localization of individual microbubbles in a frame-by-frame manner, and as such, may be suitable for real-time implementation. We demonstrate the effectiveness of the proposed approach on both simulations and {\it in-vivo} contrast enhanced human prostate scans, acquired with a clinically approved scanner.   
\end{abstract}

\begin{IEEEkeywords}
Ultrasound, Contrast agents, Super-resolution, Compressed sensing, Kalman filter.
\end{IEEEkeywords}

\section{Introduction}
In the past several decades, ultrasonic contrast agents have been utilized successfully in numerous applications \cite{schlief1991ultrasound, de1991principles, cosgrove2006ultrasound}. In particular, contrast enhanced ultrasound (CEUS) imaging takes advantage of inert microbubbles (MBs) which are injected into the bloodstream, as means to image blood vessels with improved contrast, compared with standard B-mode ultrasound (US) imaging \cite{hudson2015dynamic}. In recent years, super-resolution US imaging emerged, and enabled the fine visualization and detailed assessment of capillary blood vessels {\it in vivo} \cite{Errico2015, oreilly2013super, christensen2015vivo, ackermann2016detection, foiret2017ultrasound,bar2017fast}. Super-resolution US relies on concepts borrowed from super-resolution fluorescence microscopy techniques such as photo-activated localization microscopy (PALM) and stochastic optical reconstruction microscopy (STORM) \cite{Betzig2006, Rust2006}, which localize individual fluorescing molecules with sub-pixel precision over many frames and sum all localizations to produce a super-resolved image. In CEUS, individual resonating MBs, similar in size to red blood cells, serve as point emitters. Their subsequent localizations are then accumulated to produce the final super-resolved image of the vascular bed with a ten-fold improved spatial resolution compared with standard CEUS imaging. To produce a reliable reconstruction, low MB concentrations are typically used \cite{oreilly2013super, christensen2015vivo}, such that in each frame all MBs are well isolated from one another. The localization procedure then amounts to pin-pointing the centroid of a single Gaussian for each detected MB in the captured movie. 

Despite yielding a substantial improvement in the spatial resolution, super-resolution ultrasound imaging typically requires tens of thousands of images to produce a single super-resolved image. Acquisition of such a large number of frames results in long scanning durations, leading to poor temporal resolution on the reconstructed sequence. Furthermore, clinical bolus doses injected to human patients result in high overlap between different MBs \cite{van2017ultrasound}. These limitations hamper the clinical applicability of localization-based super-resolution techniques. 


To overcome the temporal limitation of localization-based super-resolution while not compromising the spatial resolution of the reconstructed image, sparsity-based \cite{eldar2015sampling} approaches were recently proposed \cite{bar2017sparsity, van2017sparsity, bar2017sparsity2}. These approaches favor overlapping MBs to reduce the total acquisition time. As such, sparsity-based methods achieve faster temporal resolution using standard clinical concentrations of MBs. In \cite{bar2017sparsity}, sparsity-based super-resolution ultrasound hemodynamic imaging (SUSHI), using ultrafast plane-wave acquisition, demonstrated a super-resolved time-lapse movie of 25Hz, showing super-resolved hemodynamic changes in blood flow within a rabbit's kidney. In \cite{van2017sparsity}, using a clinically approved scanner with an acquisition rate of 10Hz, a super-resolved image of a human prostate vasculature was demonstrated. In this work, a clear depiction of vascular bifurcations was obtained, although significant MB overlap was present, by performing frame-by-frame sparse localization and subsequent accumulation of all localizations to produce the final super-resolved image.

One major difference between super-resolution in ultrasound and in microscopy is that the point emitters in ultrasound are flowing inside the blood vessels, whereas in microscopy fluorescent molecules are fixed to the sub-cellular organelles. Since the motion of individual MBs is not random but rather within blood vessels, this can be exploited to improve the recovery process. 
Here, we build on our previous results on super-resolution ultrasound imaging \cite{van2017sparsity,bar2017sparsity} by exploiting the flow kinematics of individual MBs as an additional prior in the sparse recovery process. 

While previous super-resolution works focused on ultrafast plane-wave image acquisition, e.g. \cite{Errico2015, bar2017sparsity, bar2017sparsity2}, most clinically used scanners are low-rate scanners (10-25Hz). In this work, our aim is to bridge the gap between super-resolution techniques and data obtained from research platforms in laboratory environments, typically low-rate intensity images where significant MB overlap is present. By doing so, as we demonstrate in \cref{Sec:results}, our technique enables practitioners to analyze readily available CEUS scans, and obtain both architectural as well as functional blood flow information. Such analysis can expedite the process of gaining new insights regarding cancer diagnosis \cite{van2017entropy}, treatment, {\it in-vivo} flow characterization \cite{ackermann2016detection} and more.

Our method combines weighted sparse recovery with simultaneous tracking of the individual MBs in the imaging plane. MBs flow inside blood vessels, hence their movement from one frame to the next is structured. Therefore, MBs are more likely to be found in certain areas of the next frame, given their current locations. In our method, each MB track is used to estimate the position of the MBs in the next frame, and since capillary flow is non-turbulent (peak Reynolds number of $0.001$) \cite{jensen1996estimation, ackermann2016detection}, a linear propagation model is used to describe MBs flow from one frame to the next. The accumulated position estimates are then used to form a weighting matrix for weighted sparse recovery which locates the MBs. This allows to favor more likely locations in the sparse recovery process.  
With the addition of each new frame, the tracks are updated online. 
We refer to our method as simultaneous sparsity-based super-resolution and tracking, or Triple-SAT. 
Since our approach tracks individual MBs, it is possible to also estimate their velocities. We provide {\it in-vivo} super-resolution CEUS imaging of a human prostate, and show that our velocity estimation agrees with previously published results \cite{van2017ultrasound}. An illustration of our proposed concept can be seen in \cref{{Fig:Fig1}}. 

The methods proposed in this work relate to those presented in \cite{ackermann2016detection}, in which the authors incorporated an automated detection and tracking mechanism for localized MBs. 
However, Triple-SAT differs from \cite{ackermann2016detection} in the following ways. First, in \cite{ackermann2016detection} the automatic tracking algorithm is not used to improve the localization procedure over consecutive frames. Rather, individual MBs were localized over all frames with low MBs concentration, and only then detection and tracking was performed on the localizations to improve the velocities estimation. 
Here, we use detections from previous frames to improve the detections in the next frame, using sparse recovery to overcome MBs overlap, due to the use of clinical bolus doses. 
Second, we exploit coarse measurements of MB movements based on optical flow (OF) estimation \cite{barron1994performance, horn1981determining, baker2004lucas, lucas1981iterative} over the captured low-resolution sequence to improve the tracking performance. Thus, we incorporate not only position measurements but also velocity measurements in the adopted Kalman filtering framework \cite{kalman1960new, bar1995multitarget}. These measurements help in improving the overall tracking performance  of the MBs, which in turn improves the sparse recovery process. Typically, OF estimation is performed over sequential pairs of images. Here we combine OF estimation with Kalman filtering, as means to include additional information from previous frames and improve the overall estimation performance. 

The rest of the paper is organized as follows. In \cref{Sec:Algorithm} we provide a full description of our method and describe each of its building blocks. \cref{Sec:results} presents {\it in-silico} as well as {\it in-vivo} results. In \cref{Sec:discussion} we provide a discussion and conclusions.

Throughput the paper, $x$ represents a scalar, ${\bf x}$ a vector, ${\bf X}$ a matrix and ${\bf I}_{N\times N}$ is the $N\times N$ identity matrix. The notation $||\cdot||_p$ represents the standard $p$-norm and $||\cdot||_F$ is the Frobenius norm. Subscript $x_l$ denotes the $l$th element of ${\bf x}$ and ${\bf x}_l$ is the $l$th column of ${\bf X}$. Superscript ${\bf x}^{(p)}$ represents ${\bf x}$ at iteration $p$, $T^*$ denotes the adjoint of T, and ${\bf \bar{A}}$ is the complex conjugate of ${\bf A}$. The estimated vector in frame $k$, given the estimate in the $(k-1)$th frame, is indicated by $\v s_{k|k-1}$. Likewise $\v P_{k|k-1}$ indicates its estimated covariance matrix $k$, given the $k-1$ estimate. The $ij$th element of matrix $\v A$ is denoted as $A[i,j]$.

\begin{figure}[!t]
	\centerline{\includegraphics[width=0.25\textwidth]{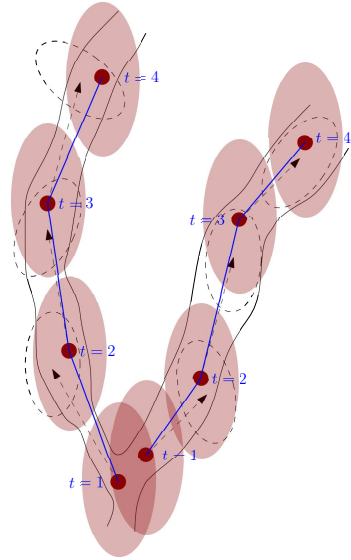}}
    \caption{Proposed concept of Triple-SAT. Individual MBs flow within a blood vessel, depicted here as a bifurcation by black solid contour lines. Large, transparent ellipses represent the echoes measured from individual MBs. In frame $t=1$ MBs are localized using sparse recovery (small red dots). Using a Kalman filter, their positions are propagated to the next frame ($t=2$), as marked by the dashed black arrows. Using the error covariance matrix of the filter, ellipses of most likely positions for the MBs are generated, as illustrated by the dashed black ellipses. These ellipses are then used as weights in the sparse recovery process in the next frame ($t=2$), and so on. The blue lines indicate the estimated trajectories of the MBs.}\label{Fig:Fig1}
\end{figure}

\section{Simultaneous sparsity-based super-resolution and tracking}
\label{Sec:Algorithm}
\subsection{Principle}
In this work, we primarily aim at improving sparsity-based super-resolution ultrasound from movies which were acquired from low-rate clinical scanners, were we have access only to the final intensity images which are displayed on screen. We start from a contrast-enhanced ultrasound sequence of $K$ frames where each frame consists of $M\times M$ pixels. A contrast-specific imaging mode based on a power-modulation pulse scheme is used to reject tissue signal and enhance signal from MBs \cite{ascenti2004contrast, bar2017fast}, such that only MBs are visualized. Prior to Triple-SAT processing, all frames are registered, as described in \cite{van2017sparsity}. After registration, we have $K$ registered low-resolution frames, which serve as the input data to our algorithm. 

\begin{figure*}[!t]
	\centerline{\includegraphics[width=0.5\textwidth]{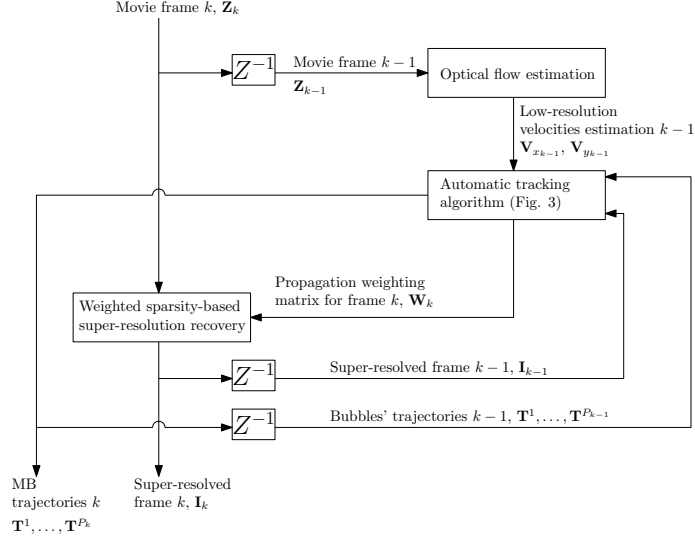}}
    \caption{Main building blocks of Triple-SAT. First, MBs velocities are estimated from frame $k$ using optical flow estimation. The localized MBs from the previous frame $k-1$ are then propagated to frame $k$, assuming a constant acceleration model using the Kalman filter and the measured velocities obtained from the OF estimation. This procedure yields a new estimate for the true MB positions and velocities, while also forms the weighting matrix for frame $k$. This weighting matrix is then used in the sparse recovery process to yield the super-resolved frame $k$. This process repeats itself for each new frame in the movie. $Z^{-1}$ represents a delay of one time-unit.}\label{Fig:Fig2}
\end{figure*}

Figure \ref{Fig:Fig2} shows the main flow and building blocks of Triple-SAT. Given the weighting matrix, based on trajectories estimated from the $(k-1)$th frame, we perform weighted sparse recovery to estimate the positions of the MBs on a high-resolution grid in the $k$th frame. Next, We acquire in the $k$th frame a crude velocity measurement of the MBs by applying OF estimation on the captured low-resolution sequence. Thus, for each MB, we obtain both positions and corresponding velocities measurements, which are then used in the automatic tracking algorithm to update the positions and velocities of the individually localized MBs using Kalman filtering. The newly estimated positions and velocities are used to generate an updated weighting matrix for sparse recovery of MB positions in the $(k+1)$th frame, while providing quantitative information on the flow kinematics.  

The reconstruction process of Triple-SAT can be considered as sparse recovery with time varying support, where the support represents the MBs locations. Previous works on sparse recovery with varying support have been proposed in the compressed sensing literature, such as \cite{vaswani2010modified, vaswani2008kalman, angelosante2009lasso, angelosante2009compressed}. Triple-SAT differs from previous works in the following manner. First, previous works assume a first-order recursion for the propagation model of the non-zero entries of the sparse signals, i.e. $x_{k+1}=\alpha x_{k} + v_{k}$, where $x_{k}$ is a scalar entry from the sparse vector, $v_{k}$ is additive Gaussian noise and $\alpha$ is a known constant. In this case, only the support of the sparse signal is of interest, but in CEUS, MBs kinematics also include varying velocities. Here we consider an extended  model which does not include only the position estimation of the MBs, but also their velocities. 
Second, as MBs flow over time, new MBs emerge and some MBs vanish from the imaging plane, due to the 3D geometry of the blood vessels. It is thus desirable to associate new MBs to previous localizations to improve the overall tracking and to achieve a more reliable estimation of their motion kinematics. This association process is not considered in prior works, but is taken into account in Triple-SAT by the use of an automatic association algorithm, combined with Kalman filtering.

We next detail the main building blocks of Triple-SAT.

\subsection{Weighted sparse recovery}
We first start with a description of our sparse recovery algorithm. This recovery procedure is performed for each frame separately. Similar to \cite{bar2017sparsity}, we model each frame as a summation of $L_k$ individual MB echoes,
\begin{equation}
\label{Eq:StreamOfPulses}
Z_k(x,z) = \sum_{i=1}^{L_k}u(x-x_i, z-z_i)\sigma_i,
\end{equation}
were $u(\cdot, \cdot)$ is the point spread function (PSF) of the transducer, $\sigma_i$ is the magnitude of the returned echo from the $i$th MB located at position $x_i,z_i$ and we assume that the PSF of the transducer is known. This PSF can be measured from the acquired data, as described in \cite{bar2017sparsity, van2017sparsity}. 

Following similar derivations to \cite{bar2017sparsity}, we discretize the $k$th frame in \cref{Eq:StreamOfPulses} as $\v Z_k, k=1,\ldots,K$ of size $M\times M$, and denote its vectorized form $\v z_k$.  We also introduce a high-resolution grid of size $N\times N$ pixels, such that $N=PM$ for some $P\geq 1$. We denote the (vectorized) super-resolved frame $k$, which contains the locations of the MBs on the high-resolution grid, by $\v i_k$. Using knowledge of the PSF, the measured frame $\v z_k$ is related to the super-resolved frame $\v i_k$ via 
\begin{equation}\label{Eq:LinModel}
{\bf z}_k={\bf H}{\bf i}_k,
\end{equation} 
where ${\bf H}$ is a known dictionary matrix based on the PSF. We follow \cite{solomon2017sparcom, bar2017sparsity} and consider recovering $\v i_k$ in the discrete Fourier domain. In this domain, $\v H$ has the following structure
\begin{equation*}
	{\bf H} = {\bf U}({\bf F}_M\otimes {\bf F}_M),
\end{equation*}
were ${\bf U}$ is an $M\times M$ diagonal matrix, whose diagonal contains the discrete Fourier transformed PSF, and $\v F_M$ is a partial $M\times N$ discrete Fourier transform (DFT) matrix, whose $M$ rows contain the $M$ lowest frequencies of a full $N\times N$ DFT matrix. Considering \cref{Eq:LinModel} in the discrete Fourier domain leads to a numerically efficient sparse solver, as described in \cite{solomon2017sparcom}. Estimation of ${\bf i}_k$ is achieved by solving the following convex optimization problem,
\begin{equation}\label{Eq:Convmin}
\min_{{\bf i}_k\geq 0}||{\bf z}_k-{\bf H}{\bf i}_k||_2^2+\lambda ||{\bf i}_k||_1,
\end{equation}
where $\lambda\geq0$ is a regularization parameter.

As in \cite{van2017sparsity}, the super-resolved image can be constructed by solving \cref{Eq:Convmin} for each frame $k$ and accumulating all localizations. To improve the sparse recovery process, we propose solving the following weighted $l_1$ minimization problem,
\begin{equation}\label{Eq:ConvminP}
\min_{{\bf i\geq 0}_k}||{\bf z}_k-{\bf H}{\bf i}_k||_2^2+\lambda ||{\bf W}_{k}{\bf i}_k||_1.
\end{equation}
The matrix $\v W_k$ is an $N^2\times N^2$ diagonal weights matrix which incorporates the flow dynamics of the MBs in the sparse recovery process, and changes with each frame. Intuitively, this matrix assigns higher weights to locations less probable to contain MBs, thus forces the sparse recovery process to favor specific locations in the frame, which are more likely to contain the MBs. In practice, we minimize \cref{Eq:ConvminP} using the FISTA algorithm \cite{Beck2009, eldar2012compressed}, as we describe in \cref{Alg:FbF_fista} or by using the reweighted iterative $l_1$ algorithm \cite{candes2008enhancing}. We apply \cref{Alg:FbF_fista} for each frame in the movie separately.

    \begin{algorithm}
        \caption{FISTA for minimizing \cref{Eq:ConvminP}}
        \label{Alg:FbF_fista}
        \begin{algorithmic}
            \REQUIRE $\v z_k$, $\v W_k$, $\lambda>0$, maximum iterations $D_{\textrm{max}}$
            \STATE {\bf Initialize} ${\bf y}_1={\bf x}_0={\bf 0}$, $\v w=\textrm{diag}\{\v W_k\}$, $t_1=1$ and $d=1$
            \WHILE {$d\leq D_{\textrm{max}}$ or stopping criteria not fulfilled}
                \STATE {\bf 1:} $\v g_d=\v H^H\v H\v y_d-\v H^H\v z_k$
                \STATE {\bf 2:} ${\bf x}_d=\Tscr_{\frac{\lambda}{L}\v w}({\bf y}_d-{\frac{1}{L}}\v g_d)$
                \STATE {\bf 3:} Project to the non-negative orthant ${\bf x}_d({\bf x}_d < {\bf 0}) = {\bf 0}$
                \STATE {\bf 4:} $t_{d+1}=\frac{1}{2}(1+\sqrt{1+4t_d^2})$
                \STATE {\bf 5:} ${\bf y}_{d+1}={\bf x}_d+{\frac{t_d-1}{t_{d+1}}}({\bf x}_d-{\bf x}_{d-1})$
                \STATE {\bf 6:} $d\leftarrow d+1$
            \ENDWHILE
            \RETURN $\v i_k={\bf x}_{D_{\text{max}}}$
        \end{algorithmic}
    \end{algorithm}

In \cref{Alg:FbF_fista}, $L_f$ is the Lipschitz constant of the quadratic term of \cref{Eq:ConvminP}, readily given as the maximum eigenvalue of $\v H^T\v H$ and $\Tscr_\alpha$ is the soft-thresholding operator, defined as
\begin{equation*}
\Tscr_{\bm \alpha}(\v x)[i]=\max\left(0,|x_i|-\alpha_i\right)\cdot\textrm{sign}(x_i),
\end{equation*}
where ${\bm \alpha}$ and $\v x$ are vectors of the same length.

We next describe how to construct $\v W_k$ per frame using MBs trajectories, Kalman filtering and OF.

\subsection{Microbubble tracking}\label{Seq:Tracking}
The weighting matrix $\v W_k$ represents the accumulated probability of localized MBs from the $(k-1)$th frame to be found in new locations in the $k$th frame. 
Its construction requires identifying and tracking individual MBs, as we show in \cref{Sec:W}.  
We now turn to explain this process. 
First, we define the state of the $p$th MB in frame $k$, as $\v s^p_k\in\bbR^4$ with 
\begin{equation*}
\v s^p_k=[x^p_k, {v_x}^p_k, y^p_k, {v_y}^p_k]^T.
\end{equation*}
Here, $x^p_k$ and $y^p_k$ are Cartesian coordinates which indicate the position of the $p$th MB in frame $k$, and ${v_x}^p_k$ and ${v_y}^p_k$ its respective velocities. 
The accumulation of all states of the $p$th MB from frame $1$ to frame $K_p$, $\v T^p=[\v s^p_1,\ldots,\v s^p_{K_p}]\in\bbR^{4\times K_p}$, is referred to as the track of the $p$th MB. 

To proceed, we consider an arbitrary frame, $k$. At this stage, we posses 
all the states of $P$ previously tracked MBs, $\v s^1_{k-1},\ldots,\v s^P_{k-1}$. 
Given the next low-resolution frame $\v z_k$, we have two goals:
\begin{enumerate}
\item Recover the locations of the $L$ MBs which are embodied in frame $\v z_k$. We note that the number $L$ of MBs in frame $k$ is generally different than the number of MBs in the previous frame $P$. This possible discrepancy occurs since blood vessels have a three-dimensional topology, and consequentially MBs may shift in and out of the imaging plane.
\item Associate each newly localized MB to a previously known track, or open a new track if no such association is possible. This association enables to produce the weighting matrix $\v W_k$ by propagating the tracks of individual MBs, while it also provides 
estimation of MB velocities.
\end{enumerate}

\begin{figure}[!t]
	\centerline{\includegraphics[width=0.5\textwidth]{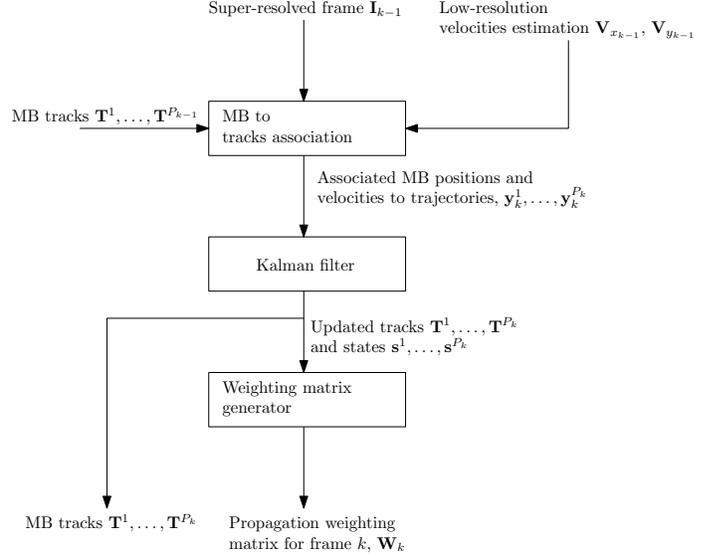}}
    \caption{Automatic tracking and data association procedure. New localized MBs from the $(k-1)$th super-resolved frame are associated to previously known $P_{k-1}$ tracks, or open new tracks, while non-associated tracks are closed (uppermost block, realized by the MHT algorithm). Then, using Kalman filtering, these new $P_k$ tracks are propagated to the next frame (central block). Estimated velocities in the xy plane ($\v V_{x_{k-1}}, \v V_{y_{k-1}}$) using optical flow on the low resolution $k-1$ frame are associated to the newly localized MBs and used as measured velocities for the Kalman filter update. Thus, an updated track estimation is produced. Lastly, these propagated tracks form the weighting matrix $\v W_k$ (lowest block). The tracks serve as inputs to the algorithm in the next frame, when new localizations arrive, and the tracking and association process repeats itself.}\label{Fig:Fig3}
\end{figure}

The tracking and association process is illustrated in \cref{Fig:Fig3}. The output of \cref{Alg:FbF_fista} is the $(k-1)$th super-resolved frame, $\v i_{k-1}$, whose non-zero values correspond to the positions of the MBs present in this frame. Next, given all previously known tracks $\v T^1,\ldots,\v T^P$, these positions need to be associated to the tracks. The updated tracks are essential to the formulation of $\v W_k$. 
The goal of the uppermost block in \cref{Fig:Fig3} is to associate each individually localized MB to one of the known $P$ tracks, or to open a new track if no such correspondence is found. Specifically, this matching and association process is realized using the multiple hypothesis tracking (MHT) procedure. 

The MHT algorithm, as first suggested by Reid \cite{reid1979algorithm, bar1995multitarget}, is considered as one of the most popular data association algorithms. The key idea in MHT is to produce a tree of potential hypotheses for each target, in our case MB locations. Upon receiving new measurements, the likelihood of each possible track is calculated and the most likely tracks are selected. This can be done by formulating and solving the maximum weighted independent set \cite{kim2015multiple, pardalos1991algorithm}, for example. The likelihood calculation relies on all past observations of each target \cite{kim2015multiple}. The MHT algorithm is known to produce good data association results due to its pruning stage. Ambiguities are assumed to be resolvable when new data is acquired. As such, given the latest measurements in frame $k$, the method estimates the likelihood based upon $J$ previous measurements (where $J$ can be controlled) to resolve past ambiguities in the $(k-J)$th frame irrevocably, and updates all tracks accordingly for the current frame. Thus, data-to-track association decisions are always based upon previous $J$ frames, in a sliding-window manner. An example of associated track numbers to new localizations is presented in \cref{Fig:Fig5}, left panel.   
In practice, we use the Lisbon implementation, taken from \cite{antunes2011library, antunes2011multiple}, which offers full integration into the MATLAB environment. 

At the end of this stage, after the association process is finished, existing tracks have been assigned new measurements (MBs positions and velocities), and new tracks are generated, if new MBs were localized. If an existing track was not updated, then this track is closed and cannot be further updated, indicating that the individual MB of this track is no longer present in the movie. We now need to integrate the measurements to their corresponding tracks, and propagate the updated tracks to the next frame $k$.  

Track update and propagation is performed by applying Kalman filtering to each track, individually. Individual tracks represent the history of each localized MB. This history helps to propagate the MBs to the next frame more accurately, and to obtain better velocity estimation. To this end, we consider the $p$th track. We assume a linear propagation model for the locations of the individual MBs between consecutive frames. This model is given by
\begin{equation}
\label{Eq:PropModel}
\v s^p_k = {\bm \Phi}\v s^p_{k-1} + {\bm \eta}^p_k, 
\end{equation}
where 
\begin{equation*}
{\bm \Phi} = \left[\begin{array}{cccccc}
1 & \Delta T & 0 & 0\\
0 & 1       & 0 & 0\\
0 & 0 		& 1 & \Delta T \\
0 & 0 		& 0 & 1
\end{array}\right],
\end{equation*}
with $1/\Delta T$ being the frame-rate of the US machine. Model \cref{Eq:PropModel} corresponds to the discretized version of the {\it continuous white noise acceleration (CWNA) model}, or {\it second-order kinematic model} \cite{bar2004estimation}. Ideally, a constant velocity model has zero acceleration, or zero second-order derivative. In practice, the CWNA model assumes that the velocity of each MB has slight perturbations, modeled as a zero-mean white noise with power spectral density $\rho$. In the discrete model \cref{Eq:PropModel}, this uncertainty is captured by the zero-mean additive Gaussian noise vector ${\bm \eta}^p_k$, associated with a covariance matrix $E\{{\bm \eta}^p_k{\bm  \eta}^{p^T}_k\}=\v Q^p_k$. Following \cite{bar2004estimation}, the CWNA model covariance matrix $\v Q^p_k$ is given by 
\begin{equation*}
\v Q^p_k = \left[\begin{array}{cccccc}
1/3\Delta T^3& 1/2\Delta T^2 & 0 & 0 \\
1/2\Delta T^2& \Delta T & 0 & 0 \\
0 & 0 & 1/3\Delta T^3& 1/2\Delta T^2 \\
0 & 0 & 1/2\Delta T^2& \Delta T \\
\end{array}\right]\rho,
\end{equation*}
where $\rho$ is chosen empirically. 

The measurement model for the $p$th MB is given by
\begin{equation}\label{Eq:MeasModel}
\v y^p_k = \v s^p_{k-1} + {\bm \zeta}^p_k, 
\end{equation}
were 
${\bm \zeta}^p_k$ is zero-mean independent i.i.d. Gaussian noise with covariance matrix $E\{{\bm \zeta}^p_k{\bm \zeta}^{p^T}_k\}=\v R^p_k$. In practice, we choose this matrix to be a diagonal matrix, and consider its diagonal as free parameters of Triple-SAT, which are chosen empirically. 

From the super-resolved image $\v i_{k-1}$ we measure the position of the MBs. 
Specifically, consider an MB which is detected in position $[n_k\Delta_X,n_y\Delta_Y]^T$, where $\Delta_X$ and $\Delta_Y$ are the known sizes of each pixel in the super-resolved image and $[n_k,n_y]$ are some integers. If the MHT algorithm decided that this MB belongs to the $p$th track, then $y_k^p[1]=n_k\Delta_X$ and $y_k^p[3]=n_k\Delta_Y$. The velocities of the MBs, or $y_k^p[2]$ and $y_k^p[4]$, are measured using OF estimation \cite{barron1994performance} on the low-resolution movie frames, as will be described in \cref{Sec:OFest}. 

Given the propagation \cref{Eq:PropModel} and the measurement \cref{Eq:MeasModel} models, we now formulate the Kalman filter update rules. The model propagation and its corresponding propagated estimation covariance matrix are given by
\begin{equation}\label{Eq:PropEq}
\begin{array}{ll}
\v s^p_{k|k-1} = {\bm \Phi} \v s^p_{k-1|k-1},\\
\v P^p_{k|k-1} = {\bm \Phi}\v P^p_{k-1|k-1} {\bm \Phi}^T + \v Q^p_k.
\end{array}
\end{equation}
Using \cref{Eq:PropEq}, the weighting matrix $\v W_k$ is calculated, as we describe in \cref{Sec:W}. Next, we solve \cref{Eq:ConvminP} to recover the $k$th super-resolved frame, $\v i_{k}$. 
After the association process is finished, for each track
we update its last state via the Kalman filter equations. The Kalman gain is given by
\begin{equation}\label{Eq:Kgain}
{\bf K}^p_k={\bf P}^p_{k|k-1}({\bf P}^p_{k|k-1}+{\bf R}^p_k)^{-1},
\end{equation}
and the innovation step along with the updated estimation error covariance matrix are given by
\begin{equation}\label{Eq:Innov}
\begin{array}{ll}
{\bf s}^p_{k|k}={\bf s}^p_{k|k-1}+{\bf K}^p_k({\bf y}^p_k-{\bf s}^p_{x|x-1})\\ 
{\bf P}^p_{k|k}=({\bf I}_{6\times 6}-{\bf K}^p_k){\bf P}^p_{x|x-1}.
\end{array}
\end{equation}
From the innovation step \cref{Eq:Innov}, we update the states as ${\bf s}^p_{k}={\bf s}^p_{k|k}$ with estimation covariance matrix ${\bf P}^p_{k}={\bf P}^p_{k|k}$.  

\subsection{Weighting matrix formulation}\label{Sec:W}
\begin{figure}[!tb]
	\centerline{\includegraphics[width=0.5\textwidth]{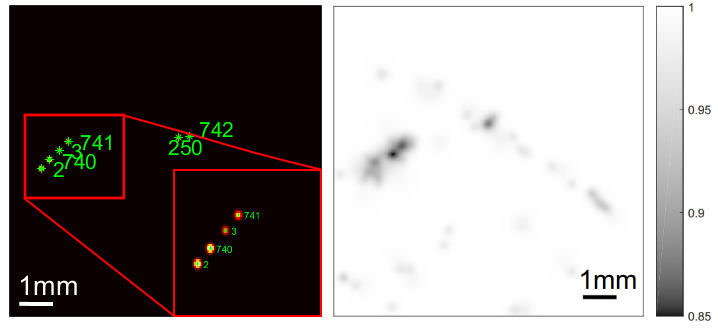}}
    \caption{Left panel shows an example of association of new measurements to existing tracks. Red box indicates enlarged area. MBs were smoothed slightly for visualization purposes only. Right panel depicts an example of the weighting matrix $\v W_k$, presented as an $N\times N$ image.}\label{Fig:Fig5}
\end{figure}

\begin{figure}[!tb]
	\centerline{\includegraphics[width=0.25\textwidth]{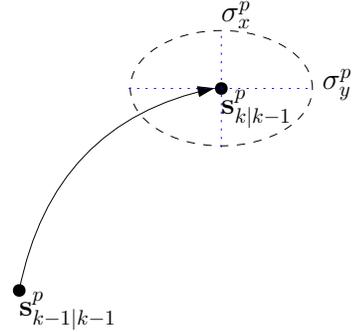}}
    \caption{Generation of the weighting matrix $\v W_k$. Previously estimated state $\v s^p_{k-1|k-1}$ is propagated to state $\v s^p_{k|k-1}$, according to \cref{Eq:PropEq}. Its propagated error covariance matrix $\v P^p_{k|k-1}$ is then used to draw an ellipse around its location, where $\sigma^p_x=\v P^p_{k|k-1}[1,1]$ and $\sigma^p_y=\v P^p_{k|k-1}[3,3]$. Aggregation of all propagated uncertainty ellipses generates an image of possible MBs locations. $\v W_k$ is then proportional to the inverse of this image.}\label{Fig:Fig4}
\end{figure}

After the states for all MBs are propagated using \cref{Eq:PropEq} and associated to existing or new tracks, we turn to formulate the weighting matrix $\v W_k$, as illustrated in \cref{Fig:Fig3}, lowest panel. The propagated state ${\bf s}^p_{k|k-1}$ represents the position and velocity of the $p$th MB, and has its associated estimation error covariance matrix $\v P^p_{k|k-1}$. 
Based on the state predictions, we formulate a spatial MB-likelihood map $\v J_k$, by assigning probabilities drawn from an anisotropic Gaussian distribution of which the mean and covariance are dictated by their respective predictions/updates in the Kalman framework. 
This process is illustrated in \cref{Fig:Fig4}. By aggregating the estimated positions and Gaussians of all of the $P$ propagated MBs, a spatial map of their possible true locations on the high-resolution grid is constructed, denoted as $\v J_k$. The $ij$th element of this $N\times N$ matrix is expressed as
\begin{equation*}
J_k[i,j] = \dsum_{p = 1}^{P}A^p e^{-q^p\left( \frac{1}{{\sigma^p_x}^2}(i - x^p_0)^2 - c^p(i-x^p_0)(j-y^p_0) + \frac{1}{{\sigma^p_y}^2}(j - y^p_0)^2 \right)},
\end{equation*}
with $A^p = \sqrt{|2\pi\v P^p_{k|k-1}|}$, $[x^p_0,y^p_0]=[s^p_{k|k-1}[1],s^p_{k|k-1}[3]]$, $\sigma^p_x=P^p_{k|k-1}[1,1]$, $\sigma^p_y=P^p_{k|k-1}[3,3]$, $q^p=1/(2(1-{\rho^p}^2))$, $c^p=2\rho^p/(\sigma^p_x\sigma^p_y)$ and $\rho^p=P^p_{k|k-1}[1,3]/(\sigma^p_x\sigma^p_y)$. The diagonal of the weighting matrix $\v W_k$ is the inverse of the vectorized form of $\v J_k$ plus a small regularization value $\epsilon$, to avoid division by zero,
\begin{equation}
\label{Eq:WeightMat}
\begin{array}{ll}
W_k[i,i] = \frac{1}{J_{k}[\floor{i/N},(i\mod N)]+\epsilon},\\
i=1,\ldots,N^2,
\end{array}
\end{equation}
 where $\floor{\cdot}$ is the floor operation and $(x\mod y)$ is the modulo operation with the swap $0\rightarrow N$. An example illustration of such a weighting matrix can be observed in \cref{Fig:Fig5}, right panel. Vectorization of this $N\times N$ image is the diagonal of $\v W_k$. The main building blocks of Triple-SAT are described in \cref{Alg:MainFlowAlg}.

As noted before, in \cref{Eq:MeasModel} we assume that we measure not only the positions of localized MBs, but also their velocities. We now turn to describe how this velocity measurement is done.

    \begin{algorithm}
        \caption{Triple-SAT}
        \label{Alg:MainFlowAlg}
        \begin{algorithmic}
            \REQUIRE Low-resolution movie $\v Z_k, k=1,\ldots,K$
            \WHILE {$k\leq K$}
                \STATE {\bf 1:} Estimate OF on $\v z_k$ using MATLAB's \matst{opticalFlow} command
                \STATE {\bf 2:} Perform sparse super-resolution on $\v z_k$ using \cref{Alg:FbF_fista}
                \STATE {\bf 3:} reconstruct new measurement vectors for all $P$ localized MBs $\v y_k^p, p=1,\ldots,P$
                \STATE {\bf 4:} Associate $\v y_k^p$ to existing tracks $\v T^p, p=1,\ldots,P$ / open new tracks / close old tracks using the MHT algorithm
                \STATE {\bf 5:} Update last state of existing / new tracks $\v T^p$ using \cref{Eq:Innov}
                \STATE {\bf 6:} Propagate last state of existing / new tracks $\v T^p$ with \cref{Eq:PropEq}
                \STATE {\bf 7:} Construct weighting matrix $\v W_{k+1}$ using \cref{Eq:WeightMat}
                \STATE {\bf 8:} $k\leftarrow k+1$
            \ENDWHILE
            \RETURN Super-resolved frames $\v i_k, i=1,\ldots,K$ and MB tracks 
        \end{algorithmic}
    \end{algorithm}

\subsection{Velocity regularization via optical flow estimation}\label{Sec:OFest}
To improve the tracking procedure of individual MBs, we provide velocity measurements to the Kalman filter as part of the input to \cref{Eq:MeasModel}. This is done by OF estimation \cite{barron1994performance, horn1981determining, baker2004lucas, lucas1981iterative} from the low-resolution movie. Although the formulation originates from the low-resolution movie, in which individual MBs are not separable, this added velocity information helps in regularizing the tracking of individual MBs. Consider for example, a newly localized MB. This MB has a single position measurement. Without added information of its general direction of movement, the tracking filter will propagate the MB to a position, which in general is not related to its actual position in the next frame. If, on the other hand, additional information in the form of its coarse velocity is supplied, then the filter will propagate the MB to a location in which the MB is more likely to be detected in the next frame. 

We now provide a short description of how OF can be estimated from the low-resolution frames, and how these measurements are incorporated in the Kalman filter framework described in \cref{Seq:Tracking} to update the state of each localized MB.

The basic assumption in OF estimation is the pixel intensity consistency assumption, or
\begin{equation}
\label{Eq:OFass}
Z_k[x,y] = Z_{k+1}[x+u_{x_k},y+u_{y_k}],
\end{equation}
where $[x,y]$ are all pixel coordinates and $[u_{x_k},u_{y_k}]$ are their corresponding two-dimensional velocities at frame $k$. Expanding $Z_{k+1}[x+u_{x_k},y+u_{y_k}]$ to its first order Taylor series around pixel $[x,y]$ yields
\begin{equation*}
Z_{k+1}[x+u_{x_k},y+u_{y_k}]\approx Z_k[x,y]+\v u_k^T{\bm \nabla} Z_k[x,y]+Z_{k_t}[x,y],
\end{equation*}
where $\v u_k=[u_{x_k},u_{y_k}]^T$, ${\bm \nabla} Z_k[x,y] = [Z_{k_x}[x,y], Z_{k_y}[x,y]]^T$ are the spatial derivatives of $Z_k[x,y]$ and $Z_{k_t}[x,y]$ is the temporal derivative of $Z_k[x,y]$ between frames $k+1$ and $k$.

Since we perform registration of the images prior to Triple-SAT, the only objects moving in the US sequence are the MBs. Hence, the OF estimation process estimates their velocities on the low resolution grid. By enforcing \cref{Eq:OFass}, we obtain the gradient constraint equation
\begin{equation}
\label{Eq:GradConst}
\v u_k^T{\bm \nabla} Z_k[x,y]=-Z_{k_t}[x,y].
\end{equation}
Denoting 
\begin{equation*}
\begin{array}{llll}
\v v_k = [u_{x_k}[1,1],\ldots,u_{x_k}[M,M], u_{y_k}[1,1],\ldots,u_{y_k}[M,M]]^T,\\
\v Z_{k_t} = -[Z_{k_t}[1,1],\ldots,Z_{k_t}[M,M]]^T,\\
\v A_{x_k}=\textrm{diag}[Z_{k_x}[1,1],\ldots,Z_{k_x}[M,M]],\\
\v A_{y_k}=\textrm{diag}[Z_{k_y}[1,1],\ldots,Z_{k_y}[M,M]],
\end{array}
\end{equation*}
with $\v D_k=[\v A_{x_k}, \v A_{y_k}]$, \cref{Eq:GradConst} is written in matrix form as 
\begin{equation}
\label{Eq:GradConstMat}
\v D_k\v v_k= \v Z_{k_t}.
\end{equation}
Equation \cref{Eq:GradConstMat} represents an ill-posed model, since $\v v_k$ has $2M^2$ unknowns, but only $M^2$ measurements are given in $\v D_k$. 

Many ways to regularize and estimate the OF exist. For example, Horn and Schunk \cite{horn1981determining} assume global $l_2$ smoothness of the OF field \cref{Eq:GradConst} over all image pixels to estimate the OF and solve
\begin{equation*}
\min_{\tilde{\v u_k}} ||\v D_k\v v_k-\v Z_{k_t}||_2^2+\mu ||{\bm \nabla} \v u_k||_2^2,
\end{equation*}
for some regularization parameter $\mu>0$, where ${\bm \nabla} \v u_k$ are the spatial derivatives of $\v u_k$. Lucas and Kanade \cite{lucas1981iterative} assume that locally, the OF field is constant and minimize a least-squares criterion,  
\begin{equation*}
\min_{\v v_k} ||\v M_k(\tilde{\v D}_k\tilde{\v v}_k-\v Z_{k_t})||_2^2,
\end{equation*}
with $\tilde{\v v}_k\in\bbR^2$ being the lateral and axial velocities,   
\begin{equation*}
\tilde{\v D}_k=
\left[\begin{array}{cc}
Z_{k_x}[1,1] & Z_{k_y}[1,1]\\
\vdots & \vdots\\
Z_{k_x}[M_\Omega,M_\Omega] & Z_{k_y}[M_\Omega,M_\Omega]
\end{array}\right],
\end{equation*}
$\v M_k$ being a window weighting matrix which favors pixels at its center compared with pixels at its periphery and $M_\Omega < M$. The latter minimization problem is solved locally for a small neighborhood on a pixel-wise based manner. OF estimation methods are easily implemented using the \matst{opticalFlow} command in MATLAB. In practice we achieved good performance with the method of Lukas and Kanade with a Gaussian smoothing kernel and a standard deviation of $1.5$ pixels. 

\begin{figure*}[!t]
    \centerline{\includegraphics[width=1\textwidth]{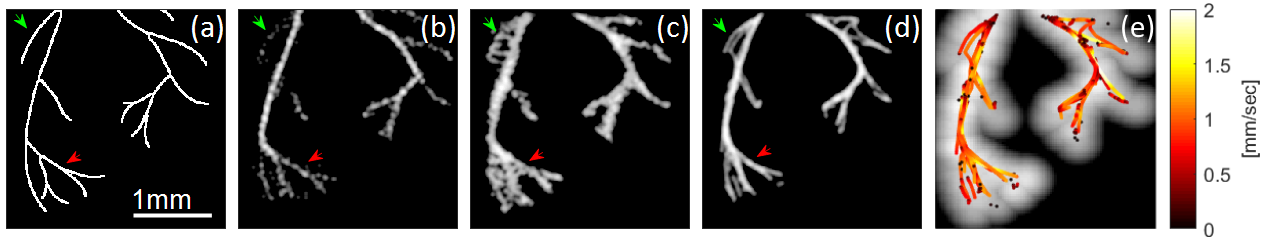}}
    \caption{Simulation results. (a) Ground truth image of bifurcating blood vessels. (b) Super-localization reconstruction. (c) Super-resolution sparse recovery obtained by minimizing \cref{Eq:Convmin} via FISTA. (d) Triple-SAT recovery, by accumulating all recovered MB trajectories. (e) Superimposed velocity trajectories over the maximum intensity projection (MIP) image, obtained from the Triple-SAT recovery. 
All reconstructions are displayed in logarithmic scale, with a dynamic range of 25dB.}\label{Fig:Sim}
\end{figure*}

In practice, each low-resolution frame $\v Z_k$ is first interpolated to the size of the $N\times N$ super-resolved images $\v I_k$, and OF estimation is performed subsequently. This procedure ensures that each pixel in the super-resolved image is associated with a velocity vector from its corresponding interpolated low-resolution frame. Together, the obtained velocities are considered as measurements for the Kalman filter, along with MBs localizations from the super-resolved frame $\v I_k$. 
Formally, we denote the $xy$ velocity fields obtained by OF estimation over the interpolated low-resolution frame $\v Z_k$ as $\v V_{x_k}$ and $\v V_{y_k}$. That is, both $\v V_{x_k}$ and $\v V_{y_k}$ are $N\times N$ matrices, and each of their pixels corresponds to the pixel-wise lateral and axial estimated velocities, respectively. Next, for MBs localized in pixels $[i_p,j_p]$ from $\v I_k$, we associate the corresponding velocity values from $\v V_{x_k}$ and $\v V_{y_k}$, 
\begin{equation}
\begin{array}{ll}
y_k^p[2] = V_{x_k}[i_p, j_p], \\
y_k^p[4] = V_{y_k}[i_p, j_p], \;\;p=1,\ldots,P.
\end{array}
\end{equation}
Thus, the first and third entries of the measurement vector $\v y^p_k$ in \cref{Eq:MeasModel} represent the measured position of the $p$th localized MB in the $k$th frame, and the second and fourth entries represent its measured velocity. 

Note, that we perform OF estimation on the low-resolution movie and not on the super-resolved frames, since we observed in practice that the pixel intensity consistency assumption \cref{Eq:OFass} does not hold on the super-resolved images. This is because a typical super-resolved image looks like the image displayed in \cref{Fig:Fig5}, left panel. The enlarged box shows the localizations of four MBs (smoothed only for display purposes). Typically, in the next frame these MBs move considerably, such that \cref{Eq:OFass} does not hold. In contrast, this assumption holds much better on the low-resolution images, due to the spreading of the diffraction-limited echoes from the MBs over several adjacent pixels.

Since the velocities are measured on the low-resolution images, from non-resolved MBs, we cannot expect them to represent the measured velocities of individually resolved MBs. Rather, these velocities represent a coarse, low-resolution estimate of the averaged velocities from the non-resolved MBs. As such, the velocity measurements are weighted with ten-times larger values in the measurement covariance matrix $\v R_k^p$ than the position measurements, to represent the inherent inaccuracy of their measurement.

\section{Materials and methods}
\label{Sec:Materials}
\subsection{Numerical simulations}
We simulated a bolus injection into vascular bifurcations over an acquisition period of $105$ frames, with frame rate of $10$Hz, such that the total acquisition time is $10.5$ seconds. Pixel size is $0.03\times 0.03\text{mm}^2$. The received modulation frequency was 7MHz (second harmonic of $3.5$MHz, similar to our {\it in-vivo} acquisition setup), with a Gaussian PSF having a standard deviation of 0.14mm in the axial direction and $0.16\text{mm}$ in the lateral direction. MB velocities' magnitudes and directions were generated by taking the maximum between values drawn from a normal distribution with a mean of $1\text{mm/sec}$ and standard deviation of $1\text{mm/sec}$, and zero. 


We set $P=3$ and recover the super-resolved images on a three times denser grid than the low-resolution grid with $\lambda=0.035$. We iterate over $4000$ iterations per frame and set $\epsilon=0.01$, $\rho=300$, and $\v R_k=\textrm{diag}\{0.1, 1, 0.1, 1\}$. In the MHT algorithm, we set the probability for not detecting an existing target as 0.1, the probability for a new target to appear as 0.1, the probability for false-alarm as 0.01 and a maximum number of leaves of 6. 

\begin{figure*}[!t]
    \centerline{\includegraphics[width=1\textwidth]{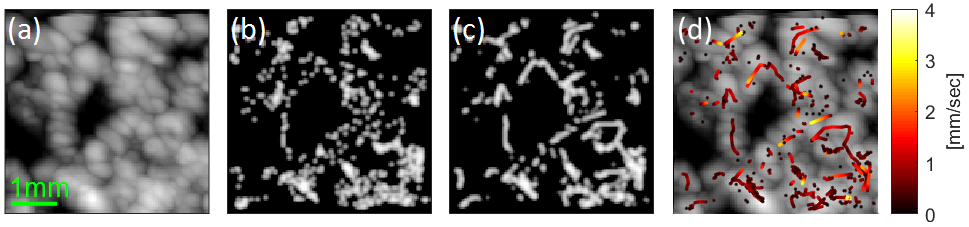}}
    \caption{Triple-SAT applied to an {\it in-vivo} scan from a human-prostate. (a) MIP image from 100 frames. (b) Super-resolution sparse recovery obtained by minimizing \cref{Eq:Convmin} via FISTA. (c) Triple-SAT recovery, by accumulating all recovered MB trajectories. (d) Superimposed velocity trajectories over the MIP image, obtained from the Triple-SAT recovery. 
All images are displayed in logarithmic scale, with a dynamic range of 30dB.}\label{Fig:Patch1}
\end{figure*}

\begin{figure*}[!t]
    \centerline{\includegraphics[width=1\textwidth]{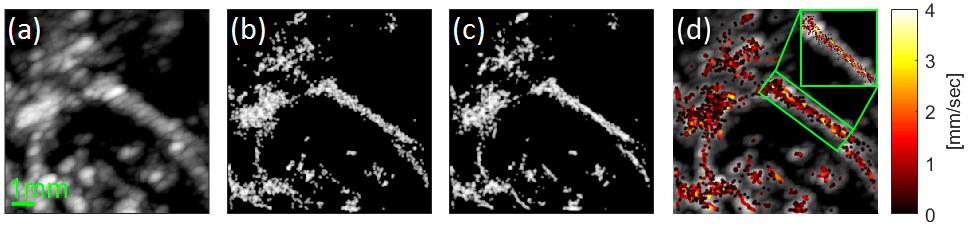}}
    \caption{Additional example of Triple-SAT recovery of an {\it in-vivo}, human-prostate scan. (a) MIP image from 150 frames. (b) Super-resolution sparse recovery obtained by minimizing \cref{Eq:Convmin} via FISTA. (c) Triple-SAT recovery by accumulating all recovered MB trajectories. (d) Superimposed velocity trajectories over the MIP image, obtained from the Triple-SAT recovery. 
All images are displayed in logarithmic scale, with a dynamic range of 30dB.}\label{Fig:Patch2}
\end{figure*}

\subsection{{\it In-vivo} experiments}
CEUS data of human prostates was acquired at the AMC university hospital (Amsterdam, the Netherlands), using a transrectal ultrasound probe and an iU22 scanner (Philips Healthcare, Bothell, WA) operating in a contrast-specific mode at a frame rate of $1/\Delta T=10\text{Hz}$. A 2.4-mL MB bolus of Sonovue\textsuperscript{TM} (Bracco, Milan, Itally) was administered intravenously, and 100-150 frames (10-15 seconds) were collected for further analysis. Pixel size is $0.146\times 0.146\text{mm}^2$.

We consider two examples taken from the {\it in-vivo} scan. For all experiments, we set $P=4$ and recover the super-resolved images on a four times denser grid than the low-resolution grid. 
In both cases we use $2000$ iterations per frame, $\epsilon=0.5$, $\rho=500$, and $\v R_k=\textrm{diag}\{0.1, 1, 0.1, 1\}$. For both examples, in the MHT algorithm, we set the probability for not detecting an existing target as 0.1, the probability for a new target to appear as 0.5, the probability for false-alarm as 0.01 and a maximum number of leaves of 6. 

\section{Results}
\label{Sec:results}
\subsection{Numerical simulations}
Figure \ref{Fig:Sim} shows reconstruction results of the simulated dataset of flowing MBs within a simulated vascular network. Panel (a) shows the ground truth architecture, while panels (a) (b) and (c) show the reconstruction results of standard super-localization \cite{ovesny2014thunderstorm}, sparsity-driven super-resolution \cite{van2017sparsity} (minimizing \cref{Eq:Convmin}) and Triple-SAT, respectively. Panel (e) shows an overlay of MB sub-diffraction trajectories, colored by their estimated velocities over the maximum intensity projection image (MIP) image. 

We observe that the Triple-SAT recovery (panel (d)) seems the smoothest and most continuous, compared with the super-localization and sparsity-based reconstructions, depicted in panels (b) and (c), respectively. 
The green arrows in the panels show a protruding blood vessel, which is almost absent in panel (b), and in panel (c) it seems connected to an additional vessel. In contrast, in panel (d), this blood vessel is more clearly depicted. The red arrows indicate clear bifurcations in the Triple-SAT reconstruction (panel d), which are absent in the sparsity-based recovery (panel c). This area is more clearly reconstructed in the super-localization image (panel d) than in the sparsity-driven recovery, but seems less smooth and clear compared with the Triple-SAT image. Though Triple-SAT shows some reconstruction artifacts, it depicts the clearest and smoothest image compared to other methods.  

Additionally, panel (e) reveals that the estimated velocities are in the range $0-2\text{mm/sec}$, as expected from the simulation. 
A histogram of the measured velocities is given in panel (c) of \cref{Fig:Hist}, where the velocities distribution is indeed between $0-2\text{mm/sec}$. 

\subsection{{\it In-vivo} experiments}
In this section, we present {\it in-vivo} reconstruction results of Triple-SAT. Ultrasound acquisition parameters and reconstruction parameters are given in \cref{Sec:Materials}. Figures \ref{Fig:Patch1} and \ref{Fig:Patch2} compare between different reconstructions in two areas of a prostate CEUS scan. In both figures, panel (a) shows the MIP image. This image is diffraction limited and was generated as reference for standard non-super-resolution image processing by taking the pixel-wise maximum value over the entire movie. Panel (b) shows sparsity-based super-resolution as obtained by minimizing \cref{Eq:Convmin} via FISTA, while panel (c) shows the Triple-SAT image. Lastly, panel (d) displays an overlay of the estimated velocities' trajectories on the MIP image.  

\begin{figure}[!t]
    \centerline{\includegraphics[trim={2.2cm 3cm 1.5cm 3.5cm},clip,width=0.54\textwidth]{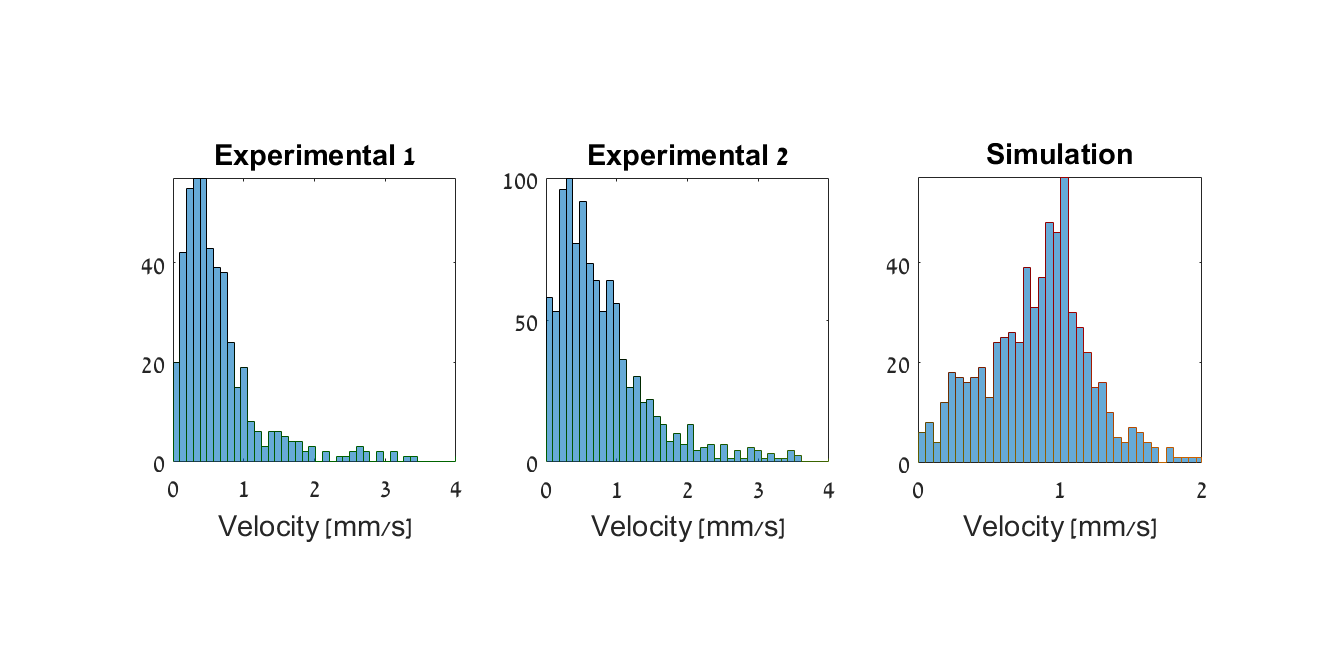}}
    \caption{Estimated velocity histograms. (a) Experimental dataset in \cref{Fig:Patch1}. (b) Experimental dataset in \cref{Fig:Patch2}. (c) Simulation dataset in \cref{Fig:Sim}.}\label{Fig:Hist}
\end{figure}

Comparing panels (a) to (b) and (c) of both figures, we notice how sparse recovery (panel (b)) achieves a super-resolved depiction of the fine vasculature. However, the Triple-SAT image (panel (c)) qualitatively appears more smooth and continuous, showing distinct trajectories which are absent in the sparse recovery images. Finally, panels (d) in both figures present velocity magnitude estimations from the tracked MBs. The plotted trajectories are sub-diffraction in size, which is clearly noticeable for example in panel (d) of \cref{Fig:Patch1}. The green box in the upper-right corner depicts an enlarged region, were clear MB trajectories with velocities of $\sim2-3\text{mm/sec}$ are observed. 
The vast majority of obtained MB flow velocities are on the order of up to $1-2\text{mm/sec}$, in line with previous observations on blood flow in micro-vessels \cite{van2017ultrasound}. This is also confirmed by the velocity magnitude histograms, displayed in \cref{Fig:Hist}. 

Panels (d) in both figures depict many point-like and short trajectories, with low velocities, alongside longer and smoother trajectories. We attribute these point-like trajectories to the fact that the vascular bed of the prostate is inherently 3D, with many blood vessels crossing the imaging plane of the probe. Thus, lateral and axial velocities (with respect to the transducer position) of MBs flowing within these blood vessels can be small. In contrast, the simulated MBs in \cref{Fig:Sim} are simulated in a plane, and clearly show long and smooth trajectories. 

\section{Discussion and conclusions}
\label{Sec:discussion}
This is the first work to exploit the inherent motion kinematics of individual MBs as a structural prior for super-resolution. Since individual MBs flow within blood vessels, their positions can be predicted from one frame to the next. In Triple-SAT we exploit this additional information to improve sparse-recovery, by solving a support aware minimization problem, as formulated in \cref{Eq:ConvminP}. Using Kalman filtering, we track and propagate the trajectories of individual MBs from one frame to the next. Moreover, we introduce velocity measurements via optical flow estimation to improve the tracking process for super-resolution imaging. Figures \ref{Fig:Patch1} and \ref{Fig:Patch2} show the power and potential of Triple-SAT on {\it in-vivo} data. Both clear and smooth super-resolution imaging are achieved, as well as a quantitative measurement of the flow velocities of individual MBs. 

As with other sparsity-based super-resolution methods, Triple-SAT operates well with high MB concentrations, for which significant MB overlap is present. This is a key observation in bridging the gap between low-rate clinical scanners with a fixed frame-rate, and clinical MBs doses. We note that Triple-SAT can also work with even higher frame-rates, such as fast plane-wave scans. This, for example, would allow to produce super-resolved images in approximately $\leq10$ms using $100-150$ frames acquired with a frame-rate of $10$KHz. By exploiting the sparse nature of the individual MB echoes, Triple-SAT is able to depict the vasculature with a relatively low number of frames. In our {\it in-vivo} experiments we used two datasets of 100 and 150 frames. However, as MB density increases even further, several mechanisms of Triple-SAT might fail. First, the sparse-recovery algorithm may not be able to accurately detect and localize all of the MBs in each frame. Second, MHT data-to-track association may also fail to properly associate new localizations to existing tracks, as resolved MBs become extremely close to one another. Lastly, OF estimation will fail to produce reliable results in areas of many overlapping MBs which move in different directions, resulting in an almost-zero averaged velocity estimate on the low-resolution grid. Reducing the frame-rate even more will also cause OF estimation and MHT data-to-track association to fail. The former, since the basic assumption of pixel intensity consistency breaks down as the frame-rate decreases. The latter, since the association of new measurements to existing tracks becomes less likely than the opening of new tracks, even if they belong to previous tracks. Yet, as we report in this work, for clinical bolus doses and $10$Hz scanners, Triple-SAT reliably recovers the vascular bed.  

In \cite{bar2017sparsity}, a similar model to \cref{Eq:StreamOfPulses} was introduced over the beam-formed complex IQ signal, while here we assume such a model over the real-valued intensity images. Nonetheless, in practice, both in simulations and {\it in-vivo} experiments, we observed good reconstruction performance using model \cref{Eq:StreamOfPulses}, as we presented in \cref{Sec:results}.

The use of a Kalman filter for MB tracking has two main motivations. First, the Kalman filter is an online estimator, which is suitable for real-time applications. Using this filter for online tracking of MBs can lead to a real-time clinical application of Triple-SAT. Second, it is known that capillary flow is non-turbulent \cite{jensen1996estimation, ackermann2016detection}. Thus, a simple linear propagation model can be used for the tracking procedure.

There are two main limitations to Triple-SAT. The first is inherent to all ultrasound super-resolution techniques. MBs must flow through the vasculature in order to detect it, thus setting a minimal acquisition time for any super-resolution imaging technique. In Triple-SAT, by using high-concentration bolus doses and sparse-recovery, we reduce the acquisition time, but only to the degree that MBs flow within the finest blood vessels during that period. Second, Triple-SAT includes several parameters which should be selected properly, among which are the sparsity regularization parameter $\lambda$, $\epsilon$, $\rho$ and the probabilities for the MHT algorithm. In this work, these parameters were chosen manually, and calibrated according to the simulation, but additional validation and testing of their selection is required.

Before concluding, we would like to discuss some computational aspects of online sparse tracking, as the number of detected MBs grows. The authors of \cite{angelosante2009lasso} suggested an $l_1$ relaxed adaptation of the Kalman filter to account for the possible exponential growth in computational complexity with the problem dimensions. In practice, although in Triple-SAT we apply a Kalman filter to each localized MB, this computational growth was not observed to be dramatic, even when tens of MBs were tracked simultaneously. We ascribe this to the fact that the state of each MB is relatively low dimensional (four entries of positions, and velocities). As such, matrix inversions are relatively inexpensive. 

Another possible computational burden stems from the MHT algorithm, which is known to grow exponentially in complexity as the number of tracks increases. In \cite{ackermann2016detection}, the authors considered a modified version of Markov Chain Monte Carlo (MCMC) data association \cite{oh2009markov} to account for this growth. The computational complexity of MHT can also be controlled by limiting the pruning depth, achieving a trade-off between accurate data association and computational complexity. In general, any automatic association algorithm may be used in the uppermost block of \cref{Fig:Fig3} instead of MHT, such as the joint probabilistic data association (JPDA) \cite{fortmann1980multi} or the MCMC algorithm of \cite{ackermann2016detection}. 

To conclude, in this work, we presented a new algorithm to improve sparsity-based super-resolution CEUS imaging. By formulating a weighted sparse recovery minimization problem, combined with on-line tracking of individual MBs, we are able to improve the sparse recovery process. With Triple-SAT we achieve a smoother depiction of the vasculature as well as provide quantitative information regarding MB kinematics. We applied our algorithm to both simulations and {\it in-vivo} human prostate scans, obtained from low frame-rate, clinically-approved US machines, demonstrating super-resolution recovery of the vascular bed with 100-150, high-MB-density frames. Since Triple-SAT employs an on-line estimation process, it may be suitable for real-time applications within commercially available US machines.

\def\IEEEbibitemsep{0pt plus .5pt}
\small
\bibliographystyle{myIEEEtran}
\bibliography{Bib_Mendeley}

\begin{thebibliography}{10}
\providecommand{\url}[1]{#1}
\csname url@samestyle\endcsname
\providecommand{\newblock}{\relax}
\providecommand{\bibinfo}[2]{#2}
\providecommand{\BIBentrySTDinterwordspacing}{\spaceskip=0pt\relax}
\providecommand{\BIBentryALTinterwordstretchfactor}{4}
\providecommand{\BIBentryALTinterwordspacing}{\spaceskip=\fontdimen2\font plus
\BIBentryALTinterwordstretchfactor\fontdimen3\font minus
  \fontdimen4\font\relax}
\providecommand{\BIBforeignlanguage}[2]{{%
\expandafter\ifx\csname l@#1\endcsname\relax
\typeout{** WARNING: IEEEtran.bst: No hyphenation pattern has been}%
\typeout{** loaded for the language `#1'. Using the pattern for}%
\typeout{** the default language instead.}%
\else
\language=\csname l@#1\endcsname
\fi
#2}}
\providecommand{\BIBdecl}{\relax}
\BIBdecl

\bibitem{schlief1991ultrasound}
R.~Schlief, ``Ultrasound contrast agents.'' \emph{Current opinion in
  radiology}, vol.~3, no.~2, pp. 198--207, 1991.

\bibitem{de1991principles}
N.~De~Jong, F.~Ten~Cate, C.~Lancee, J.~Roelandt, and N.~Bom, ``Principles and
  recent developments in ultrasound contrast agents,'' \emph{Ultrasonics},
  vol.~29, no.~4, pp. 324--330, 1991.

\bibitem{cosgrove2006ultrasound}
D.~Cosgrove, ``Ultrasound contrast agents: an overview,'' \emph{European
  journal of radiology}, vol.~60, no.~3, pp. 324--330, 2006.

\bibitem{hudson2015dynamic}
J.~M. Hudson, R.~Williams, C.~Tremblay-Darveau, P.~S. Sheeran, L.~Milot, G.~A.
  Bjarnason, and P.~N. Burns, ``Dynamic contrast enhanced ultrasound for
  therapy monitoring,'' \emph{European journal of radiology}, vol.~84, no.~9,
  pp. 1650--1657, 2015.

\bibitem{Errico2015}
\BIBentryALTinterwordspacing
C.~Errico, J.~Pierre, S.~Pezet, Y.~Desailly, Z.~Lenkei, O.~Couture, and
  M.~Tanter, ``{Ultrafast ultrasound localization microscopy for deep
  super-resolution vascular imaging},'' \emph{Nature}, vol. 527, no. 7579, pp.
  499--502, 2015.
\BIBentrySTDinterwordspacing

\bibitem{oreilly2013super}
M.~A. OˈReilly and K.~Hynynen, ``A super-resolution ultrasound method for
  brain vascular mapping,'' \emph{Medical physics}, vol.~40, no.~11, 2013.

\bibitem{christensen2015vivo}
K.~Christensen-Jeffries, R.~J. Browning, M.-X. Tang, C.~Dunsby, and R.~J.
  Eckersley, ``In vivo acoustic super-resolution and super-resolved velocity
  mapping using microbubbles,'' \emph{IEEE transactions on medical imaging},
  vol.~34, no.~2, pp. 433--440, 2015.

\bibitem{ackermann2016detection}
D.~Ackermann and G.~Schmitz, ``Detection and tracking of multiple microbubbles
  in ultrasound b-mode images,'' \emph{IEEE transactions on ultrasonics,
  ferroelectrics, and frequency control}, vol.~63, no.~1, pp. 72--82, 2016.

\bibitem{foiret2017ultrasound}
J.~Foiret, H.~Zhang, T.~Ilovitsh, L.~Mahakian, S.~Tam, and K.~W. Ferrara,
  ``Ultrasound localization microscopy to image and assess microvasculature in
  a rat kidney,'' \emph{Scientific reports}, vol.~7, no.~1, p. 13662, 2017.

\bibitem{bar2017fast}
A.~Bar-Zion, C.~Tremblay-Darveau, O.~Solomon, D.~Adam, and Y.~C. Eldar, ``Fast
  vascular ultrasound imaging with enhanced spatial resolution and background
  rejection,'' \emph{IEEE transactions on medical imaging}, vol.~36, no.~1, pp.
  169--180, 2017.

\bibitem{Betzig2006}
E.~Betzig, G.~H. Patterson, R.~Sougrat, O.~W. Lindwasser, S.~Olenych, J.~S.
  Bonifacino, M.~W. Davidson, J.~Lippincott-Schwartz, and H.~F. Hess,
  ``{Imaging intracellular fluorescent proteins at nanometer resolution.}''
  \emph{Science}, vol. 313, no. 5793, pp. 1642--1645, 2006.

\bibitem{Rust2006}
M.~J. Rust, M.~Bates, and X.~Zhuang, ``{Sub-diffraction-limit imaging by
  stochastic optical reconstruction microscopy (STORM).}'' \emph{Nature
  methods}, vol.~3, no.~10, pp. 793--795, 2006.

\bibitem{van2017ultrasound}
R.~J. van Sloun, L.~Demi, A.~W. Postema, J.~J. de~la Rosette, H.~Wijkstra, and
  M.~Mischi, ``Ultrasound-contrast-agent dispersion and velocity imaging for
  prostate cancer localization,'' \emph{Medical image analysis}, vol.~35, pp.
  610--619, 2017.

\bibitem{eldar2015sampling}
Y.~C. Eldar, \emph{Sampling Theory: Beyond Bandlimited Systems}.\hskip 1em plus
  0.5em minus 0.4em\relax Cambridge University Press, 2015.

\bibitem{bar2017sparsity}
A.~Bar-Zion, O.~Solomon, C.~Tremblay-Darveau, D.~Adam, and Y.~C. Eldar,
  ``Sparsity-based ultrasound super-resolution hemodynamic imaging,''
  \emph{arXiv preprint:1712.00648}, 2017.

\bibitem{van2017sparsity}
R.~J. van Sloun, O.~Solomon, Y.~C. Eldar, H.~Wijkstra, and M.~Mischi,
  ``Sparsity-driven super-resolution in clinical contrast-enhanced
  ultrasound,'' in \emph{Ultrasonics Symposium (IUS), 2017 IEEE
  International}.\hskip 1em plus 0.5em minus 0.4em\relax IEEE, 2017, pp. 1--4.

\bibitem{bar2017sparsity2}
A.~Bar-Zion, O.~Solomon, C.~Tremblay-Darveau, D.~Adam, and Y.~C. Eldar,
  ``Sparsity-based ultrasound super-resolution imaging,'' in \emph{Proceedings
  of the 23rd European symposium on Ultrasound Contrast Imaging}, 2017, pp.
  156--157.

\bibitem{van2017entropy}
R.~J. van Sloun, L.~Demi, A.~W. Postema, J.~J. De~La~Rosette, H.~Wijkstra, and
  M.~Mischi, ``Entropy of ultrasound-contrast-agent velocity fields for
  angiogenesis imaging in prostate cancer,'' \emph{IEEE transactions on medical
  imaging}, vol.~36, no.~3, pp. 826--837, 2017.

\bibitem{jensen1996estimation}
J.~A. Jensen, \emph{Estimation of blood velocities using ultrasound: a signal
  processing approach}.\hskip 1em plus 0.5em minus 0.4em\relax Cambridge
  University Press, 1996.

\bibitem{barron1994performance}
J.~L. Barron, D.~J. Fleet, and S.~S. Beauchemin, ``Performance of optical flow
  techniques,'' \emph{International journal of computer vision}, vol.~12,
  no.~1, pp. 43--77, 1994.

\bibitem{horn1981determining}
B.~K. Horn and B.~G. Schunck, ``Determining optical flow,'' \emph{Artificial
  intelligence}, vol.~17, no. 1-3, pp. 185--203, 1981.

\bibitem{baker2004lucas}
S.~Baker and I.~Matthews, ``Lucas-kanade 20 years on: A unifying framework,''
  \emph{International journal of computer vision}, vol.~56, no.~3, pp.
  221--255, 2004.

\bibitem{lucas1981iterative}
B.~D. Lucas, T.~Kanade \emph{et~al.}, ``An iterative image registration
  technique with an application to stereo vision,'' 1981.

\bibitem{kalman1960new}
R.~E. Kalman \emph{et~al.}, ``A new approach to linear filtering and prediction
  problems,'' \emph{Journal of basic Engineering}, vol.~82, no.~1, pp. 35--45,
  1960.

\bibitem{bar1995multitarget}
Y.~Bar-Shalom and X.-R. Li, \emph{Multitarget-multisensor tracking: principles
  and techniques}.\hskip 1em plus 0.5em minus 0.4em\relax YBs London, UK:,
  1995, vol.~19.

\bibitem{ascenti2004contrast}
G.~Ascenti, M.~Gaeta, C.~Magno, S.~Mazziotti, A.~Blandino, D.~Melloni, and
  G.~Zimbaro, ``Contrast-enhanced second-harmonic sonography in the detection
  of pseudocapsule in renal cell carcinoma,'' \emph{American Journal of
  Roentgenology}, vol. 182, no.~6, pp. 1525--1530, 2004.

\bibitem{vaswani2010modified}
N.~Vaswani and W.~Lu, ``Modified-cs: Modifying compressive sensing for problems
  with partially known support,'' \emph{IEEE Transactions on Signal
  Processing}, vol.~58, no.~9, pp. 4595--4607, 2010.

\bibitem{vaswani2008kalman}
N.~Vaswani, ``Kalman filtered compressed sensing,'' in \emph{Image Processing,
  2008. ICIP 2008. 15th IEEE International Conference on}.\hskip 1em plus 0.5em
  minus 0.4em\relax IEEE, 2008, pp. 893--896.

\bibitem{angelosante2009lasso}
D.~Angelosante, S.~I. Roumeliotis, and G.~B. Giannakis, ``Lasso-kalman smoother
  for tracking sparse signals,'' in \emph{Signals, Systems and Computers, 2009
  Conference Record of the Forty-Third Asilomar Conference on}.\hskip 1em plus
  0.5em minus 0.4em\relax IEEE, 2009, pp. 181--185.

\bibitem{angelosante2009compressed}
D.~Angelosante, G.~B. Giannakis, and E.~Grossi, ``Compressed sensing of
  time-varying signals,'' in \emph{Digital Signal Processing, 2009 16th
  International Conference on}.\hskip 1em plus 0.5em minus 0.4em\relax IEEE,
  2009, pp. 1--8.

\bibitem{solomon2017sparcom}
O.~Solomon, Y.~C. Eldar, M.~Mutzafi, and M.~Segev, ``Sparcom: Sparsity based
  super-resolution correlation microscopy,'' \emph{arXiv preprint:1707.09255}, 2017.

\bibitem{Beck2009}
A.~Beck and M.~Teboulle, ``{A Fast Iterative Shrinkage-Thresholding
  Algorithm},'' \emph{SIAM Journal on Imaging Sciences}, vol.~2, no.~1, pp.
  183--202, 2009.

\bibitem{eldar2012compressed}
Y.~C. Eldar and G.~Kutyniok, \emph{Compressed sensing: theory and
  applications}.\hskip 1em plus 0.5em minus 0.4em\relax Cambridge University
  Press, 2012.

\bibitem{candes2008enhancing}
E.~J. Candes, M.~B. Wakin, and S.~P. Boyd, ``Enhancing sparsity by reweighted
  ℓ 1 minimization,'' \emph{Journal of Fourier analysis and applications},
  vol.~14, no. 5-6, pp. 877--905, 2008.

\bibitem{reid1979algorithm}
D.~Reid, ``An algorithm for tracking multiple targets,'' \emph{IEEE
  transactions on Automatic Control}, vol.~24, no.~6, pp. 843--854, 1979.

\bibitem{kim2015multiple}
C.~Kim, F.~Li, A.~Ciptadi, and J.~M. Rehg, ``Multiple hypothesis tracking
  revisited,'' in \emph{Proceedings of the IEEE International Conference on
  Computer Vision}, 2015, pp. 4696--4704.

\bibitem{pardalos1991algorithm}
P.~M. Pardalos and N.~Desai, ``An algorithm for finding a maximum weighted
  independent set in an arbitrary graph,'' \emph{International Journal of
  Computer Mathematics}, vol.~38, no. 3-4, pp. 163--175, 1991.

\bibitem{antunes2011library}
D.~M. Antunes, D.~M. de~Matos, and J.~Gaspar, ``A library for implementing the
  multiple hypothesis tracking algorithm,'' \emph{arXiv preprint:1106.2263}, 2011.

\bibitem{antunes2011multiple}
D.~M. Antunes, D.~Figueira, D.~M. Matos, A.~Bernardino, and J.~Gaspar,
  ``Multiple hypothesis tracking in camera networks,'' in \emph{Computer Vision
  Workshops (ICCV Workshops), 2011 IEEE International Conference on}.\hskip 1em
  plus 0.5em minus 0.4em\relax IEEE, 2011, pp. 367--374.

\bibitem{bar2004estimation}
Y.~Bar-Shalom, X.~R. Li, and T.~Kirubarajan, \emph{Estimation with applications
  to tracking and navigation: theory algorithms and software}.\hskip 1em plus
  0.5em minus 0.4em\relax John Wiley \& Sons, 2004.

\bibitem{ovesny2014thunderstorm}
M.~Ovesn{\`y}, P.~K{\v{r}}{\'\i}{\v{z}}ek, J.~Borkovec, Z.~{\v{S}}vindrych, and
  G.~M. Hagen, ``Thunderstorm: a comprehensive imagej plug-in for palm and
  storm data analysis and super-resolution imaging,'' \emph{Bioinformatics},
  vol.~30, no.~16, pp. 2389--2390, 2014.

\bibitem{oh2009markov}
S.~Oh, S.~Russell, and S.~Sastry, ``Markov chain monte carlo data association
  for multi-target tracking,'' \emph{IEEE Transactions on Automatic Control},
  vol.~54, no.~3, pp. 481--497, 2009.

\bibitem{fortmann1980multi}
T.~E. Fortmann, Y.~Bar-Shalom, and M.~Scheffe, ``Multi-target tracking using
  joint probabilistic data association,'' in \emph{Decision and Control
  including the Symposium on Adaptive Processes, 1980 19th IEEE Conference
  on}.\hskip 1em plus 0.5em minus 0.4em\relax IEEE, 1980, pp. 807--812.

\end{thebibliography}
\end{document}